\documentclass[aps,twocolumn,showpacs]{revtex4}%
\usepackage{amsmath}
\usepackage{graphicx}
\usepackage{dcolumn}
\usepackage{verbatim}
\usepackage{times}
\usepackage{subfigure}
\usepackage{bm}
\usepackage{color}
\usepackage[colorlinks,dvipdfm]{hyperref}
\usepackage{subfigure}
\usepackage{amsfonts}
\usepackage{amssymb}%
\providecommand{\U}[1]{\protect\rule{.1in}{.1in}}

\begin{document}
\title{Pairing symmetry of heavy fermion superconductivity in the two-dimensional
Kondo-Heisenberg lattice model}
\author{Yu Liu$^{1}$, Guang-Ming Zhang$^{2,3,**}$, and Lu Yu$^{1,3}$}
\affiliation{$^{1}$Institute of Physics, Chinese Academy of Sciences, Beijing 100190, China.}
\affiliation{$^{2}$State Key Laboratory of Low-Dimensional Quantum Physics and Department
of Physics, Tsinghua University, Beijing 100084, China;}
\affiliation{$^{3}$Collaborative Innovation Center of Quantum Matter, Beijing, China}
\affiliation{$^{**}$Correspondence author. Email: gmzhang@tsinghua.edu.cn}
\date{\today }

\begin{abstract}
In the two-dimensional Kondo-Heisenberg lattice model away from half-filled,
the local antiferromagnetic exchange coupling can provide the pairing
mechanism of quasiparticles via the Kondo screening effect, leading to the
heavy fermion superconductivity. We find that the pairing symmetry
\textit{strongly} depends on the Fermi surface (FS) structure in the normal
metallic state. When $J_{H}/J_{K}$ is very small, the FS is a small hole-like
circle around the corner of the Brillouin zone, and the s-wave pairing
symmetry has a lower ground state energy. For the intermediate coupling values
of $J_{H}/J_{K}$, the extended s-wave pairing symmetry gives the favored
ground state. However, when $J_{H}/J_{K}$ is larger than a critical value, the
FS transforms into four small hole pockets crossing the boundary of the
magnetic Brillouin zone, and the d-wave pairing symmetry becomes more
favorable. In that regime, the resulting superconducting state is
characterized by either nodal d-wave or nodeless d-wave state, depending on
the conduction electron filling factor as well. A continuous phase transition
exists between these two states. This result may be related to the phase
transition of the nodal d-wave state to a fully gapped state, which is
recently observed in Yb doped CeCoIn$_{5}$.

\end{abstract}

\pacs{71.27.+a, 74.70.Tx, 75.30.Mb}
\maketitle

The study of heavy fermion intermetallic compounds has played an important
role in our understanding of strongly correlated electron
systems\cite{Stewart,Lohneysen}, and the Kondo lattice model is believed to
capture the low temperature physics\cite{Si-2001}. It has been established
that the coherent superposition of individual Kondo screening clouds gives
rise to a huge mass enhancement of quasiparticles, leading to a heavy Fermi
liquid with a \textit{large} Fermi surface comprising conduction electrons as
well as local moments\cite{Ogata-2007,Assaad-2008}. Competing with the Kondo
singlet formation, the local moments indirectly interact with each other via
the Ruderman-Kittel-Kasuya-Yosida (RKKY) interaction, and the
antiferromagnetic (AFM) long-range ordered phase emerges in the small Kondo
exchange coupling regime\cite{Doniach,Lacroix-Cyrot,zhang-2000}.

In addition, there has been growing evidence that local AFM correlation of the
local moments can also induce the formation of Cooper pairs of heavy
quasiparticles, leading to unconventional heavy fermion superconductivity. Due
to the tiny energy gap, the direct experimental measurement of superconducting
gap function is extremely difficult. Only recently, inelastic neutron
scattering experiments on CeCu$_{2}$Si$_{2}$ have revealed a spin resonance
peak, an indirect evidence of the nodal $d$-wave
superconductivity\cite{Stockert-2011}. Thermal conductivity and heat capacity
measurements on CeCoIn$_{5}$ also supported a superconducting gap with nodes
along the diagonal directions of the Brillouin zone\cite{Izawa,Aoki-2004}.
More recently, in scanning tunneling spectroscopy experiments the nodal points
in the superconducting gap of CeCoIn$_{5}$ were found, as a more direct
evidence of d$_{x^{2}-y^{2}}$ symmetry\cite{Davis,Yazdani}.

At the same time, an s-wave superconductivity has also been confirmed in
CeRu$_{2}$ and CeCo$_{2}$ in nuclear quadrupole resonance
measurements\cite{Matsuda-1995,Ishida-1997}, where the spin-lattice relaxation
rate exhibits an exponential decay at low temperatures and the Hebel-Slichter
peak is observed. Furthermore, the laser photoemission spectroscopy
measurements on CeRu$_{2}$ have demonstrated a clear anisotropic s-wave
superconducting gap at the Fermi level\cite{Kiss-2005}. Since the Coulomb
repulsion is dominant in heavy fermion systems, these results cast doubts on
whether this fully gapped superconducting state can be understood within the
pairing mechanism of the local AFM correlation.

From the theoretical side, to simplify the RKKY interaction, one can
explicitly introduce the local AFM Heisenberg superexchange interaction
$J_{H}$ among the local moments into the Kondo lattice
system\cite{Coleman-1989,Coqblin-1997,Senthil-2004,Coleman-2005,Pepin-2008,Senthil-2009,zhang-2011}%
. When the local moments are expressed in terms of fermionic spinons, the
large-$N$ fermionic approach\cite{read-newns,Millis-1986,Auerbach-1986} can be
used to treat the Kondo-Heisenberg lattice model very efficiently in the limit
of $J_{K}>J_{H}$, after a spinon hopping order parameter is
introduced\cite{Coleman-1989,Coqblin-1997,Pepin-2008,zhang-2011}. Then an
effective hybridization between the conduction electron and spinon bands leads
to a paramagnetic heavy Fermi liquid. Based on the reconstructed large Fermi
surface (FS), the instabilities of AFM order and unconventional
superconductivity can be further analyzed. As long as the AFM long range order
is suppressed, the spinon singlet pairings can be also promoted from the local
AFM spin exchange, further reducing the ground state energy. Via the Kondo
screening effect, the Cooper pairs of the conduction electrons can be induced,
leading to heavy fermion superconductivity. Some numerical evidence on robust
d-wave pairings has been observed in the two-dimensional Kondo-Heisenberg
lattice systems\cite{Sato-2001,Dagotto,Becca-2014}.

In this paper, we develop an effective mean field (MF) theory on heavy fermion
superconductivity in the Kondo-Heisenberg lattice model on a two-dimensional
square lattice. In the paramagnetic Fermi liquid phase ($J_{K}>J_{H}$), we
first notice that the FS topology undergoes a dramatic change as the ratio of
$x=J_{H}/J_{K}$ is gradually increased. Since the spinon singlet pairing from
the local AFM exchange coupling has a form factor, which is either an extended
s-wave or d-wave symmetry, we find that the superconducting pairing symmetry
depends on the FS structure. In the presence of the spinon pairings, we have
to introduce the local pairing order parameter between the conduction
electrons and spinons in the Kondo spin exchange interaction. When $x$ is very
small, the FS is a small hole-like circle around the corner of the first
Brillouin zone. We find that the s-wave superconducting state has a lower
ground state energy\cite{Pruschke-2013}. For the intermediate coupling values
of $J_{H}/J_{K}$, the extended s-wave pairing symmetry gives the favored
ground state. However, as $x$ is larger than a critical value, the
corresponding FS consists of four hole pockets crossing the boundary of the
magnetic Brillouin zone, and the d-wave pairing symmetry is more favorable. In
this regime, the resulting superconducting state can be a nodal d-wave or
nodeless d-wave state, depending on the conduction electron filling factor as
well. This result may be used to explain the recent experimental observation
in the Yb doped CeCoIn$_{5}$ (Ref.\cite{Prozorov}).

The model Hamiltonian of the Kondo-Heisenberg lattice model is defined by:
\begin{equation}
H=\sum_{\mathbf{k},\sigma}\epsilon_{\mathbf{k}}c_{\mathbf{k}\sigma}^{\dagger
}c_{\mathbf{k}\sigma}+J_{K}\sum_{i}\mathbf{S}_{i}\cdot\mathbf{s}_{i}+J_{H}%
\sum_{\left\langle ij\right\rangle }\mathbf{S}_{i}\cdot\mathbf{S}_{j},
\end{equation}
where $\epsilon_{\mathbf{k}}$ denotes the conduction electron band, the local
moments have the fermionic representation $\mathbf{S}_{i}=\frac{1}{2}%
\sum_{\sigma\sigma^{\prime}}f_{i\sigma}^{\dagger}\mathbf{\tau}_{\sigma
\sigma^{\prime}}f_{i\sigma^{\prime}}$ and $\mathbf{\tau}$ is the Pauli
matrices. There is a local constraint: $\sum_{\sigma}f_{i\sigma}^{\dagger
}f_{i\sigma}=1$ to restrict any charge fluctuations, and the f-fermions only
describe the spin degrees of freedom of the local moments, which will be
referred to as spinons.

Following the large-$N$ fermionic approach\cite{Coleman-1989,Senthil-2004},
the Kondo spin exchange and Heisenberg superexchange terms can be simply
expressed as
\begin{align}
\mathbf{S}_{i}\cdot\mathbf{s}_{i}  &  =-\frac{1}{2}\left(  f_{i\uparrow
}^{\dagger}c_{i\uparrow}+f_{i\downarrow}^{\dagger}c_{i\downarrow}\right)
\left(  c_{i\uparrow}^{\dagger}f_{i\uparrow}+c_{i\downarrow}^{\dagger
}f_{i\downarrow}\right)  ,\nonumber\\
\mathbf{S}_{i}\cdot\mathbf{S}_{j}  &  =-\frac{1}{2}(f_{i\uparrow}^{\dagger
}f_{j\uparrow}+f_{i\downarrow}^{\dagger}f_{j\downarrow})(f_{j\uparrow
}^{\dagger}f_{i\uparrow}+f_{j\downarrow}^{\dagger}f_{i\downarrow}),
\end{align}
where constant terms have been neglected. Then a Kondo screening and a uniform
short-range AFM order parameters can be introduced as
\begin{equation}
V=-\sum_{\sigma}\left\langle c_{i\sigma}^{\dagger}f_{i\sigma}\right\rangle
,\chi=-\sum_{\sigma}\left\langle f_{i\sigma}^{\dagger}f_{j\sigma}\right\rangle
. \label{eqn2}%
\end{equation}
To avoid the accidental degeneracy of the conduction electrons on a square
lattice, we choose
\begin{equation}
\epsilon_{\mathbf{k}}=-2t\left(  \cos k_{x}+\cos k_{y}\right)  +4t^{\prime
}\cos k_{x}\cos k_{y}-\mu,
\end{equation}
where $t$ and $t^{\prime}$ are the first and second nearest neighbor hopping
parameters, and a chemical potential of the conduction electrons $\mu$ has
been introduced as a Lagrangian multiplier to fix the density of the
conduction electrons $n_{c}$. Under the uniform MF approximation, the spinons
also form a very narrow band with the dispersion $\chi_{\mathbf{k}}=J_{H}%
\chi\left(  \cos k_{x}+\cos k_{y}\right)  +\lambda$ where $\lambda$ is also a
Lagrangian multiplier to impose the local constraint on average.

Then the MF Hamiltonian for the heavy Fermi liquid phase reads
\begin{equation}
\mathcal{H}=\sum_{\mathbf{k}\sigma}\left(
\begin{array}
[c]{cc}%
c_{\mathbf{k}\sigma}^{\dagger} & f_{\mathbf{k}\sigma}^{\dagger}%
\end{array}
\right)  \left(
\begin{array}
[c]{cc}%
\epsilon_{\mathbf{k}} & \frac{1}{2}J_{K}V\\
\frac{1}{2}J_{K}V & \chi_{\mathbf{k}}%
\end{array}
\right)  \left(
\begin{array}
[c]{c}%
c_{\mathbf{k}\sigma}\\
f_{\mathbf{k}\sigma}%
\end{array}
\right)  +E_{0},
\end{equation}
with $E_{0}=N\left(  J_{H}\chi^{2}+J_{K}V^{2}/2-\lambda+\mu n_{c}\right)  $.
The quasiparticle excitation spectra can be easily obtained
\begin{equation}
E_{\mathbf{k}}^{\left(  \pm\right)  }=\frac{1}{2}\left[  \left(
\epsilon_{\mathbf{k}}+\chi_{\mathbf{k}}\right)  \pm\sqrt{\left(
\epsilon_{\mathbf{k}}-\chi_{\mathbf{k}}\right)  ^{2}+\left(  J_{K}V\right)
^{2}}\right]  . \label{eqn3}%
\end{equation}
Then the self-consistent equations for $\chi$, $V$, $\lambda$ and $\mu$ can be
derived by minimizing the ground state energy. We first perform a numerical
calculation, choosing the following parameters $t^{\prime}=0.3t$, $J_{K}=2t$,
and $n_{c}=0.8$. The following results are obtained: as the local AFM
interaction $J_{H}$ gradually increases, the lower branch of the heavy
quasiparticle spectrum $E_{\mathbf{k}}^{\left(  -\right)  }$ is calculated and
the corresponding FS topology obtained and displayed in Fig.1.
\begin{figure}[ptb]
\centering \includegraphics[scale=0.35]{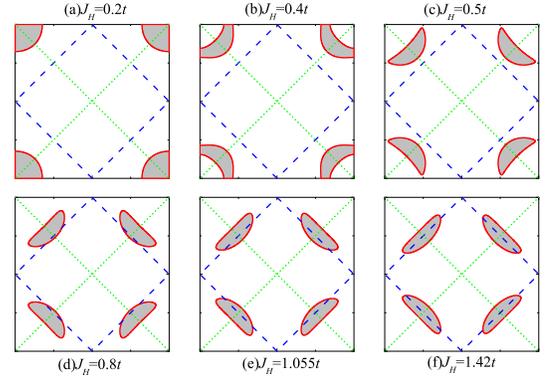}\caption{(Color online) The
Fermi surface reconstruction of the heavy quasiparticles as a function of
local AFM spin interaction $J_{H}/t$ for $J_{K}=2t$, $t^{\prime}/t=0.3$ and
$n_{c}=0.8$.}%
\end{figure}

For a very small local AFM Heisenberg spin exchange, i.e. $J_{H}/t\leq0.355$,
we can clearly see that the FS is a hole-like circle around the corner of the
first Brillouin zone, which corresponds to the large electronic FS of the
heavy quasiparticles. At $J_{H}/t=0.356$, the topology of the FS starts to
change: a small circle emerges in the center of the deformed hole FS. As
$J_{H}/t$ is further increased, both circles expand and the small one is
deformed into a rotated square. Up to $J_{H}/t=0.481$, the two deformed
circles intersect each other and then decompose into four kidney-like Fermi
pockets. When $J_{H}/t$ continues to increase, the resulting FS is shifted
inward along the diagonal direction. The detailed discussion had been
presented in our previous publication\cite{zhang-2011}. At $J_{H}/t=1.0556$,
the Fermi pockets are close to the momentum ($\pi/2,\pi/2$) and its equivalent
points. Actually such a FS structure starts to cross the boundary of the
magnetic Brillouin zone. Quantum Monte Carlo cluster approach has been used to
study the evolution of the FS close to the magnetic transition in the
two-dimensional Kondo lattice system\cite{Assaad-2008}, where the heavy
quasiparticle bands drop below the FS giving rise to hole pockets around
$\mathbf{k}=(\pi/2,\pi/2)$ and equivalent points. Our obtained FS structure is
consistent with this numerical result.

To further consider the instability of the heavy Fermi liquid state, we should
notice that the local AFM Heisenberg superexchange can also be written in
terms of the spinon singlet pairs up to a constant
\begin{equation}
\mathbf{S}_{i}\cdot\mathbf{S}_{j}=-\frac{1}{2}(f_{i{\uparrow}}^{\dag
}f_{j\downarrow}^{\dag}-f_{i\downarrow}^{\dag}f_{j{\uparrow}}^{\dag}%
)({{f}_{j\downarrow}{f}_{i{\uparrow}}-{f}_{j{\uparrow}}{f}_{i\downarrow})}.
\end{equation}
Actually, Coleman and Andrei\cite{Coleman-1989} had emphasized that the local
SU(2) gauge invariance of the local Heisenberg spin operator generally
requires the consideration of both spinon hopping and pairing order
parameters. Further arguments can be made by using the symplectic
representation of the local magnetic spins\cite{Coleman-2010}. Then the spinon
pairing parameter is introduced as%
\begin{equation}
\Delta_{ij}=-\langle f_{i{\uparrow}}^{\dag}f_{j\downarrow}^{\dag
}-f_{i\downarrow}^{\dag}f_{j{\uparrow}}^{\dag}\rangle.
\end{equation}
For a two-dimensional square lattice model, the spinon pairing order parameter
$\Delta_{ij}$ has a form factor with either an extended s-wave or the d-wave
symmetry\cite{Kotliar-Liu}. Both the extended s-wave and d-wave form factors
in the momentum space have sign change in the Brillouin zone and are shown in
Fig.2. \begin{figure}[ptb]
\centering \includegraphics[scale=0.3]{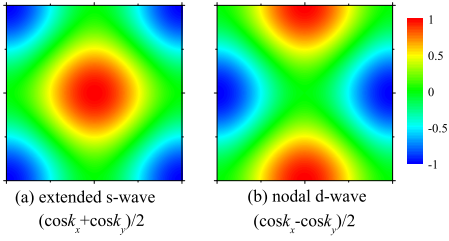}\caption{(Color
online) The extended s-wave and d-wave form factors of the spinon pairings.}%
\end{figure}

When the spinon pairings are present, the local pairing order parameter
between the conduction electrons and spinons has to be introduced, because the
Kondo spin exchange interaction can also be expressed as%
\begin{equation}
\mathbf{S}_{i}\cdot\mathbf{s}_{i}=-\frac{1}{2}\left(  c_{i\uparrow}^{\dagger
}f_{i\downarrow}^{\dagger}-c_{i\downarrow}^{\dagger}f_{i\uparrow}^{\dagger
}\right)  \left(  f_{i\downarrow}c_{i\uparrow}-f_{i\uparrow}c_{i\downarrow
}\right)  .
\end{equation}
Then a local s-wave pairing order parameter is defined by%
\begin{equation}
\Delta_{cf}=-\langle c_{i\uparrow}^{\dagger}f_{i\downarrow}^{\dagger
}-c_{i\downarrow}^{\dagger}f_{i\uparrow}^{\dagger}\rangle,
\end{equation}
so the MF model Hamiltonian in momentum space can be written in a compact
form
\begin{align}
\mathcal{H}{_{mf}}  &  =\sum_{\mathbf{k}}\psi_{\mathbf{k}}^{\dagger}\left(
\begin{array}
[c]{cccc}%
\epsilon_{\mathbf{k}} & 0 & \frac{J_{K}V}{2} & \frac{J_{K}\Delta_{cf}}{2}\\
0 & -\epsilon_{\mathbf{k}} & \frac{J_{K}\Delta_{cf}}{2} & -\frac{J_{K}V}{2}\\
\frac{J_{K}V}{2} & \frac{J_{K}\Delta_{cf}}{2} & \chi_{\mathbf{k}} &
J_{H}\Delta_{\mathbf{k}}\\
\frac{J_{K}\Delta_{cf}}{2} & -\frac{J_{K}V}{2} & J_{H}\Delta_{\mathbf{k}} &
-\chi_{\mathbf{k}}%
\end{array}
\right)  \psi_{\mathbf{k}}\nonumber\\
&  +\sum_{\mathbf{k}}\left(  \epsilon_{\mathbf{k}}+\chi_{\mathbf{k}}\right)
+E_{0},
\end{align}
where a Nambu spinor has been defined as $\psi_{\mathbf{k}}^{\dagger}=\left(
\begin{array}
[c]{cccc}%
c_{\mathbf{k}\uparrow}^{\dag} & c_{-\mathbf{k}\downarrow} & f_{\mathbf{k}%
\uparrow}^{\dag} & f_{-\mathbf{k}\downarrow}%
\end{array}
\right)  $, and $E_{0}=N\left[  J_{K}\left(  V^{2}+\Delta_{cf}^{2}\right)
/2+J_{H}\left(  \Delta_{0}^{2}+\chi^{2}\right)  -\lambda+\mu n_{c}\right]  $.
The spinon pairing gap function is chosen as
\begin{equation}
\Delta_{\mathbf{k}}=\Delta_{\mathbf{0}}(\cos k_{x}\pm\cos k_{y}),
\end{equation}
for extended s-wave and d-wave pairing, respectively. Diagonalizing this MF
model Hamiltonian, two quasiparticle bands are derived {\small
\begin{align*}
E_{\mathbf{k}}^{\pm}  &  \equiv\sqrt{E_{\mathbf{k}1}\pm\sqrt{E_{\mathbf{k}%
1}^{2}-E_{\mathbf{k}2}^{2}}},\\
E_{\mathbf{k}1}  &  \equiv\frac{1}{2}\left(  \epsilon_{\mathbf{k}}^{2}%
+\chi_{\mathbf{k}}^{2}+J_{H}^{2}\Delta_{\mathbf{k}}^{2}\right)  +J_{K}%
^{2}\left(  V^{2}+\Delta_{cf}^{2}\right)  /4,\\
E_{\mathbf{k}2}  &  \equiv\sqrt{\left[  \epsilon_{\mathbf{k}}\chi_{\mathbf{k}%
}-J_{K}^{2}\left(  V^{2}-\Delta_{cf}^{2}\right)  /4\right]  ^{2}%
+(\epsilon_{\mathbf{k}}J_{H}\Delta_{\mathbf{k}}-J_{K}^{2}V\Delta_{cf}/2)^{2}}.
\end{align*}
} Due to the particle-hole symmetry of the superconducting quasiparticles, all
negative energy states are filled up in the ground state, and the ground state
energy density is thus obtained {\small
\begin{align*}
E_{g}  &  =-\frac{\sqrt{2}}{N}\sum_{\mathbf{k}}\sqrt{E_{\mathbf{k}%
1}+E_{\mathbf{k}2}}\\
&  +\frac{J_{K}}{2}\left(  V^{2}+\Delta_{cf}^{2}\right)  +J_{H}\Delta_{0}%
^{2}+J_{H}\chi^{2}+\mu(n_{c}-1).
\end{align*}
} The saddle point equations for the MF order parameters $V$, $\chi$,
$\Delta_{0}$, $\Delta_{cf}$ and $\lambda$ can also be determined by minimizing
the ground state energy. The chemical potential $\mu$ is still determined by
the conduction electron density $n_{c}$. It should be emphasized that the
obtained MF order parameters $V$, $\chi$, and $\lambda$ are different from
those values in the normal paramagnetic phase. In particular, the difference
of the chemical potential $\mu$ modifies the position of the Fermi energy as
well as the FS structure.

Although there are no direct attractions among the conduction electrons, the
spinon singlet pairings and the pairings of the conduction electrons and
spinons provide an indirect glue for the formation of the Cooper pairs of
conduction electrons via the Kondo screening/hybridizing effect. So the
resulting ground state represents a heavy fermion superconducting state. With
the help of the double-time retarded Green function, the Cooper pairing order
parameter of the conduction electrons can also be deduced
\begin{equation}
\left\langle c_{\mathbf{k}\uparrow}^{\dag}c_{-\mathbf{k}\downarrow}^{\dag
}\right\rangle =\frac{J_{H}J_{K}^{2}\Delta_{\mathbf{k}}\left(  \Delta_{cf}%
^{2}-V^{2}\right)  +2J_{K}^{2}\chi_{\mathbf{k}}V\Delta_{cf}}{8E_{\mathbf{k}%
2}\sqrt{2\left(  E_{\mathbf{k}1}+E_{\mathbf{k}2}\right)  }}.
\end{equation}

We find that the coexisting cases ($\Delta_{cf}\Delta_{0}\neq0$) of the local
electron-spinon and spinon-spinon pairing are unstable. Therefore, only three
cases of the s-wave ($\Delta_{cf}\neq0$), the extended s-wave and d-wave
spinon pairing symmetries are discussed in the following. When we choose
$t^{\prime}=0.3t$, $n_{c}=0.8$ and $J_{K}=2t$, the MF self-consistent
equations are numerically solved, respectively. The obtained pairing strengths
are displayed in Fig.3a. When $0<J_{H}/t<1.02$, the s-wave pairing strength is
the largest. For $1.02<J_{H}/t<1.36$, the extended s-wave pairing strength has
a larger value than the d-wave pairing strength. Only for $J_{H}/t>1.36$, the
d-wave pairing strength becomes larger. Moreover, the superconducting
condensation energies are compared, we find that the s-wave pairing has a
lower ground state energy in the range $0<J_{H}/t<0.886$, the extended s-wave
pairing symmetry has a relatively lower energy in the range $0.886<J_{H}%
/t<1.056$, and the d-wave paring state becomes favorable for $J_{H}/t>1.056$.
The corresponding results are displayed in Fig.3b. As a reference, the ground
state energy of the heavy Fermi liquid phase has been subtracted to obtain the
condensation energies. As we expected, the s-wave, extended s-wave, and d-wave
symmetric superconducting states can further save the ground state energy.
\begin{figure}[ptb]
\centering \includegraphics[scale=0.3]{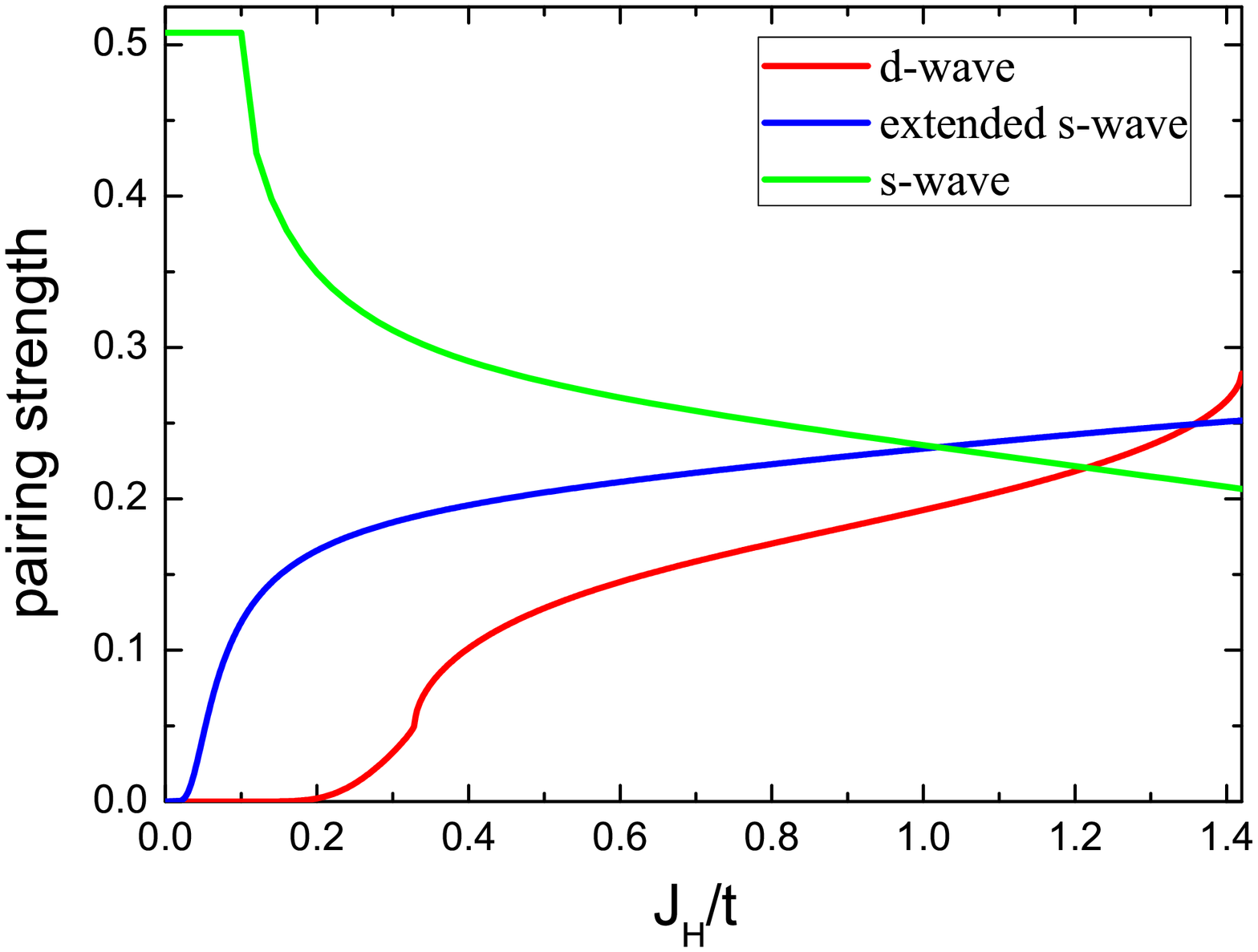}\newline%
\centering \includegraphics[scale=0.3]{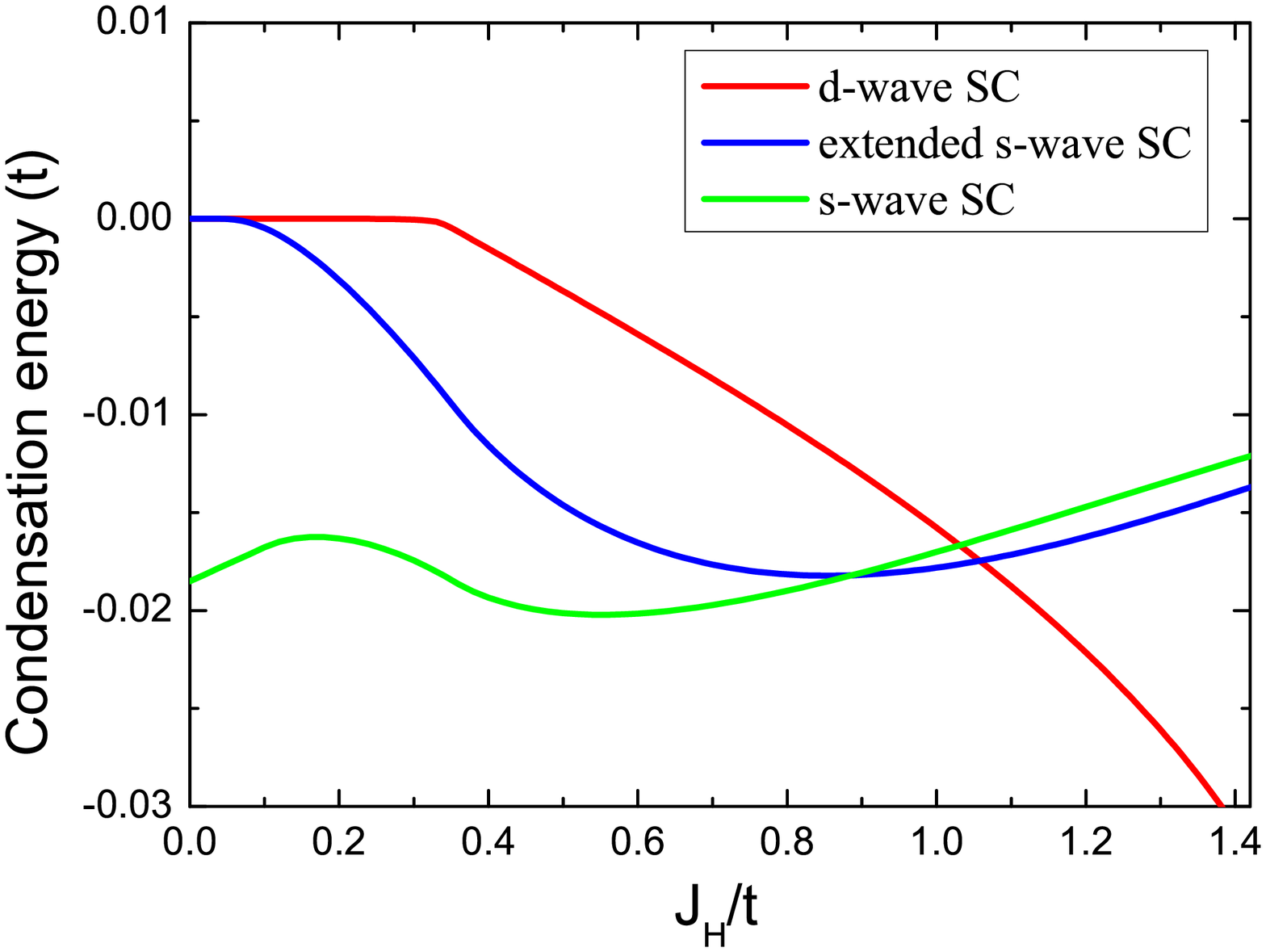}\newline%
\caption{(Color online) (Top) The pairing strengths $\Delta_{cf}$ and
$\Delta_{0}$ as a function of the local AFM spin interaction $J_{H}/t$.
(Bottom) The condensation energies with s-wave, extended s-wave, and d-wave
pairing symmetries. The other parameters are choosen as $t^{\prime}/t=0.3$,
$n_{c}=0.8$, and $J_{K}/t=2$.}%
\end{figure}

According to the above results, two discontinuous phase transition are
expected in the superconducting phase. When the local AFM spin exchange
interaction $J_{H}/t<0.886$, the s-wave pairing state is the ground state. For
$0.886<J_{H}/t<1.056$, the pairing form factor with s-wave pairing symmetry
matches the FS structure of the heavy quasiparticles, and the extended s-wave
superconducting state is the ground state of the system. However, when
$J_{H}/t>1.055$, the pairing form factor with d-wave symmetry matches the FS,
and the d-wave superconducting state becomes more favorable. Therefore, due to
the presence of the FS deformation, the resulting superconducting pairing
symmetry is very sensitive to the local AFM spin exchange.

Moreover, as the filling factor of the conduction electrons is increased,
another continuous transition can be exhibited from the lower superconducting
quasiparticle excitation spectrum. The d-wave symmetric superconducting state
shows a phase transition from the nodal d-wave to the nodeless d-wave states.
This is a very unusual phenomenon in the superconducting phase. When the Fermi
level of the normal phase is very close to the top band edge, the d-wave
pairing amplitude becomes larger than the bandwidth of the heavy quasiholes,
and the superconducting pairing state actually belongs to the strong pairing
regime. Then the nodes in the superconducting quasiparticle excitation
spectrum are no longer protected by the d-wave symmetry. In Fig.4, we show
that the lower superconducting quasiparticle excitation spectrum with
$t^{\prime}/t=0.3$, $J_{H}/t=1.42$ and $J_{K}/t=2.0$ for different conduction
electron filling factor. For the cases of $n_{c}=0.8$ and $0.829$, we can see
the nodes around ($\pi/2,\pi/2 $) and its equivalent points, while for
$n_{c}=0.87$ and $0.9$, a small gap opens up. A general ground state phase
diagram has been summarized in Fig.5. \begin{figure}[ptb]
\centering \includegraphics[scale=0.4]{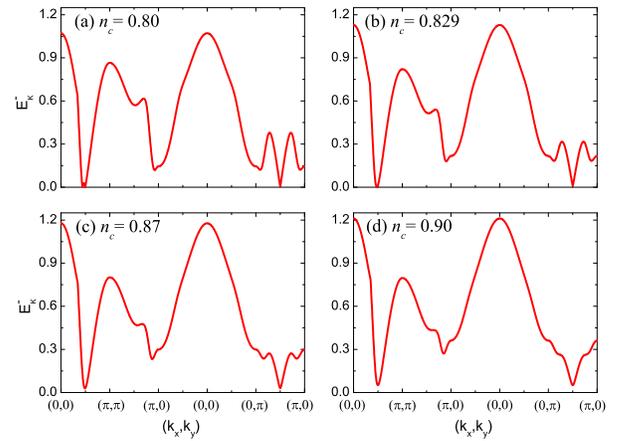}\caption{(Color online) The
lower branch of the superconducting quasiparticle excitation energy for
different conduction electron filling factor with $t^{\prime}/t=0.3$,
$J_{H}/t=1.42$ and $J_{K}/t=2.0$.}%
\end{figure}
\begin{figure}[ptb]
{\small \centering \includegraphics[scale=0.35]{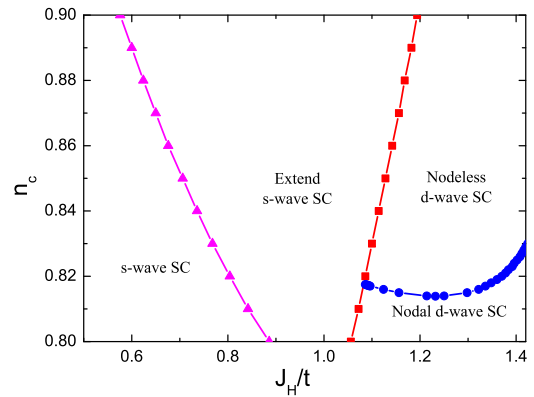}
}\caption{(Color online) The ground state phase diagram of the superconducting
state with $t^{\prime}/t=0.3$ and $J_{K}/t=2.0$.}%
\end{figure}

In conclusion, we have presented an effective MF theory for heavy fermion
superconductivity in the two-dimensional Kondo lattice model with the local
AFM Heisenberg exchange coupling among the local moments. We would like to
emphasize that, it is the local AFM short-range interaction that can deform
the FS structure of the heavy quasiparticles and induce unconventional
superconducting long-range ordered states. Due to the presence of spinon
singlet pairing, the Cooper pairs among the conduction electrons are induced
via the Kondo screening effect. However, the pairing symmetry depends on the
FS topology of heavy quasiparticles in the normal state, which is determined
by the strength of the local AFM spin exchange and the conduction electron
filling factor.

When the local AFM Heisenberg exchange coupling is weak, the heavy Fermi
liquid phase has a small hole Fermi pocket around the corner of the Brillouin
zone, and the s-wave and the extended s-wave form factor of the local AFM
interaction matches the FS, so the s-wave superconductivity is more favorable.
Such a situation occurs in the materials of CeRu$_{2}$ and CeCo$_{2}$, as
supported by various experimental
measurements\cite{Matsuda-1995,Ishida-1997,Kiss-2005}. Since these materials
show a heavy Fermi liquid behavior in the normal state, the local AFM spin
exchange interaction is expected to be very weak but can still provide an
attractive glue among the fermionic spinons.

When the local AFM Heisenberg exchange coupling is stronger than a critical
value and the conduction electron filling factor is not close to the
half-filled limit, the corresponding FS of the heavy Fermi liquid changes into
four small hole pockets around the momentum ($\pi/2,\pi/2$) and its equivalent
points. Such a FS structure more favors the d-wave form factor of the local
AFM correlation. So the nodal d-wave superconducting state can be obtained,
and such a situation happens in the materials of CeCu$_{2}$Si$_{2}$ and
CeCoIn$_{5}$, with support from various experimental
measurements\cite{Stockert-2011,Izawa,Davis}.

More recently, some interesting experimental results have been discovered upon
Yb doping in CeCoIn$_{5}$. The superconductivity is extremely stable and the
transition temperature linearly depends on the Yb doping
concentration\cite{shu-prl}, and the temperature dependence of the London
penetration depth indicates that the nodal d-wave superconductivity changes
into a fully gapped state after a critical Yb doping\cite{Prozorov}. The
transport measurements\cite{Singh-2014} have indicated that the large amount
of Yb doping introduces holes into the system. This is equivalent to
increasing the conduction electron filling factor in our theory, so the
transition observed experimentally just corresponds to the continuous phase
transition from the nodal d-wave state to the nodeless d-wave state. In order
to fully explain this transition, further theoretical work including the
estimation of the fluctuations around the MF solution should be considered.

The authors would like to thank D. H. Lee and T. Xiang for stimulating
discussions. This work was supported by the National Natural Science
Foundation of China (Grant Nos. 20121302227, 11120101003, 11121063) and by
China Postdoctoral Science Foundation (Grant No. 2013M541069).

\end{document}